\documentclass[aps,prl,twocolumn,amssymb,nofootinbib]{revtex4}
\usepackage{graphicx}
\usepackage{amsmath}
\usepackage{amssymb}
\usepackage{bm}

\usepackage{epsfig}
\usepackage{bbm}
\usepackage{color}
\usepackage{ulem}

\definecolor{MyDarkBlue}{rgb}{0.15,0.15,0.45}
\usepackage[linktocpage=true]{hyperref}
\hypersetup{
colorlinks=true,
citecolor=MyDarkBlue,
linkcolor=MyDarkBlue,
urlcolor=MyDarkBlue,
pdfauthor={Garrett Goon, Kurt Hinterbichler and Mark Trodden},
pdftitle={A New Class of Effective Field Theories from Embedded Branes},
pdfsubject={hep-th}
}

\usepackage{enumitem}
\usepackage{graphicx}
\usepackage{subfigure}
\usepackage{amsmath}
\usepackage{bbm}
\usepackage{color}

\newcommand{\be}{\begin{equation}}
\newcommand{\ee}{\end{equation}}
\newcommand{\bea}{\begin{eqnarray}}
\newcommand{\eea}{\end{eqnarray}}
\newcommand{\beas}{\begin{eqnarray*}}
\newcommand{\eeas}{\end{eqnarray*}}
\newcommand{\nn}{\nonumber}

\def\({\left(}
\def\){\right)}

\newcommand{\half}{\frac{1}{2}}

\begin{document}

\title{A New Class of Effective Field Theories from Embedded Branes} 
\author{Garrett L. Goon}
\author{Kurt Hinterbichler}
\author{Mark Trodden}

\affiliation{Center for Particle Cosmology, Department of Physics and Astronomy, University of Pennsylvania,
Philadelphia, Pennsylvania 19104, USA}
\date{\today}

\begin{abstract}
We present a new general class of four-dimensional effective field theories with interesting global symmetry groups. These theories
arise from purely gravitational actions for $3+1$-dimensional branes embedded in higher dimensional spaces with induced gravity terms. The simplest example is the well known Galileon theory, with its associated Galilean symmetry, arising as the limit of a DGP brane world. However, we demonstrate that this is a special case of a much wider range of theories, with varying structures, but with the same attractive features such as second order equations.  In some circumstances, these new effective field theories allow potentials for the scalar fields on curved space, with small masses protected by non-linear symmetries.  Such models may prove relevant to the cosmology of both the early and late universe.
\end{abstract}

\maketitle
Over the past decade, considerable interest has been paid to the induced gravity model of Dvali, Gabadadze and Poratti (DGP)~\cite{Dvali:2000hr}, and its potential applications to cosmology. This interest has been particularly strong in recent years, due to the realization that the construction of a Minkowski brane floating in a flat $5$d ambient space gives rise to a $4$d effective field theory description with interesting and nontrivial symmetry properties \cite{Luty:2003vm,Nicolis:2004qq}. The relevant symmetry, inherited from a combination of five dimensional Poincar\'e invariance and brane reparametrization invariance, takes a rather simple form in the small-field limit, and has been called the {\it Galilean} symmetry. The field obeying the symmetry describes the bending of the brane in the extra dimensional spacetime, and it has been abstracted and generalized away from DGP and referred to as the {\it Galileon}~\cite{Nicolis:2008in}. 

Among the interesting features of this Galileon field theory are that it is described by ghost-free higher derivative Lagrangians, possesses a nonrenormalization theorem~\cite{Luty:2003vm,Hinterbichler:2010xn,Burrage:2010cu}, and has possible applications to cosmology~\cite{Burrage:2010cu,Creminelli:2010ba,Creminelli:2010qf,DeFelice:2010as,Deffayet:2010qz,Kobayashi:2011pc,Mota:2010bs,Agarwal:2011mg}.  Perhaps most importantly, the theory contains a finite number of terms, and there exist regimes of interest in which these terms dominate over the infinite number of other higher order terms expected to be present in the effective field theory of a brane~\cite{Endlich:2010zj,Hinterbichler:2010xn,Nicolis:2004qq}.

In this letter, we generalize the construction through which the Galileon and Dirac-Born-infeld (DBI) theories arise from actions on a 3-brane probing a higher dimensional bulk~\cite{deRham:2010eu}. This allows us to construct entirely new classes of $4$d effective field theories, with their own interesting symmetries when the bulk possesses isometries. These theories naturally live on $4$d curved space, can retain the same number of nonlinear shift symmetries as the flat-space Galileons or DBI theories, and possess second order equations of motion.  We describe the general procedure and demonstrate important novel features such as the existence of potentials and masses fixed by symmetries, opening up the door for applications in cosmology and particle physics. In a companion paper~\cite{Goon:2011qf} we provide a detailed study of these new theories, and construct all explicit examples in the cases of maximal symmetry in both the bulk and on the brane. 


The general context of this letter is the theory of a dynamical 3-brane moving in a fixed $(4+1)$-dimensional background.  
The dynamical variables are the brane embedding $X^A(x)$, five functions of the world-volume coordinates $x^\mu$.  The bulk has a fixed background metric $G_{AB}(X)$, from which we may construct the induced metric $\bar g_{\mu\nu}(x)$ and the extrinsic curvature $K_{\mu\nu}(x)$
\bea 
\bar g_{\mu\nu}&=&e^A_{\ \mu}e^B_{\ \nu} G_{AB}(X) \ , \\ 
K_{\mu\nu}&=&e^A_{\ \mu}e^B_{\ \nu}\nabla_A n_B \ ,
\eea
where $e^A_{\ \mu}= {\partial X^A\over\partial x^\mu}$ are the tangent vectors to the brane, and $n^A$ is the unit normalized normal vector.

The world-volume action must be gauge invariant under reparametrizations of the brane,
\be 
\label{gaugetransformations} 
\delta_g X^A=\xi^\mu\partial_\mu X^A \ ,
\ee
where $\xi^\mu(x)$ is the gauge parameter.  This holds if the action is written as a diffeomorphism scalar, $F$, of $\bar g_{\mu\nu}$, $K_{\mu\nu}$, $\bar\nabla_\mu$ and the curvature $\bar R^\alpha_{\ \beta\mu\nu}$ constructed from $\bar g_{\mu\nu}$,
\be
\label{generalaction} 
S= \int d^4x\ \sqrt{-\bar g}F\left(\bar g_{\mu\nu},\bar\nabla_\mu,\bar R^{\alpha}_{\ \beta\mu\nu},K_{\mu\nu}\right) \ .
\ee

This action possesses global symmetries only if the bulk metric possesses Killing symmetries.  If the bulk metric has a Killing vector $K^A(X)$, then there is a global symmetry of the action under which the $X^A$ shift by
\be
\label{generalsym} 
\delta_K X^A=K^A(X) \ .
\ee

We fix all the gauge symmetry by choosing the gauge
\be
\label{physgauge} 
X^\mu(x)=x^\mu, \ \ \ X^5(x)\equiv \pi(x) \ ,
\ee
where the bulk is foliated by time-like slices given by the surfaces $X^5(x)= {\rm constant}$.  The remaining coordinates $X^\mu$ can then be chosen arbitrarily and parametrize the leaves of the foliation. In this gauge, the coordinate $\pi(x)$ measures the transverse position of the brane relative to the foliation, and the resulting action solely describes $\pi$,
\be
\label{gaugefixedaction} 
S= \int d^4x\ \left. \sqrt{-\bar g}F\left(\bar g_{\mu\nu},\bar\nabla_\mu,\bar R^{\alpha}_{\ \beta\mu\nu},K_{\mu\nu}\right)\right|_{X^\mu=x^\mu,\ X^5=\pi} \ .
\ee

Since gauge fixing cannot alter global symmetries, any global symmetries of~(\ref{generalaction}) become global symmetries of~(\ref{gaugefixedaction}).   However, the form of the global symmetries depends on the gauge because the gauge choice is not generally preserved by the global symmetry.  Given a transformation generated by a Killing vector, $K^A$, we restore our preferred gauge (\ref{physgauge}) by making a compensating gauge transformation $\delta_{g,{\rm comp}}x^\mu=-K^\mu$.  The two symmetries then combine to shift $\pi$ by
\be
\label{gaugefixsym} 
(\delta_K+\delta_{g,{\rm comp}})\pi=-K^\mu(x,\pi)\partial_\mu\pi+K^5(x,\pi) \ ,
\ee
which is a symmetry of the gauge fixed action~(\ref{gaugefixedaction}).


It is convenient at this stage to make two simplifying assumptions. First, we specialize to the case in which the foliation is Gaussian normal with respect to the metric $G_{AB}$.  Second, we demand that the extrinsic curvature on each of the slices be proportional to the induced metric.  Under these assumptions the metric takes the form
\be 
\label{metricform} 
G_{AB}dX^AdX^B=d\rho^2+f(\rho)^2g_{\mu\nu}(x)dx^\mu dx^\nu \ ,
\ee
where $X^5=\rho$ denotes the transverse coordinate, and $g_{\mu\nu}(x)$ is an arbitrary brane metric.  This special case includes all examples in which a maximally symmetric ambient space is foliated by maximally symmetric slices, as explored in~\cite{Goon:2011qf}.

In the gauge (\ref{physgauge}), the induced metric is
\be 
\bar g_{\mu\nu}=f(\pi)^2g_{\mu\nu}+\nabla_\mu\pi\nabla_\nu\pi \ .
\ee
Defining the quantity $\gamma=1/ \sqrt{1+{1\over f^2}(\nabla\pi)^2}$ ,
the square root of the determinant and the inverse metric may then be expressed as
\be 
\sqrt{-\bar g}=\sqrt{-g}f^4{1\over \gamma}, \ \ \ \bar g^{\mu\nu}={1\over f^2}\left(g^{\mu\nu}-\gamma^2{\nabla^\mu\pi\nabla^\nu\pi\over f^2}\right) \ .
\ee
The tangent vectors are
\be 
e^A_{\ \mu}={\partial X^A\over \partial x^\mu}=\begin{cases}\delta^\nu_\mu & A=\nu \\ \nabla_\mu \pi & A=5\end{cases} \ ,
\ee
and the normal vector $n^A$ is
\be 
n^A=\begin{cases} -{1\over f^2}\gamma\nabla^\mu\pi  \\ \gamma \end{cases},\ \ \ \ n_A=\begin{cases} -\gamma\nabla_\mu\pi & A=\mu \\ \gamma & A=5\end{cases} \ .
\ee

The extrinsic curvature is
\be 
K_{\mu\nu}=\gamma\left(-\nabla_\mu\nabla_\nu\pi+f f'g_{\mu\nu}+2{f'\over f}\nabla_\mu\pi\nabla_\nu\pi\right) \ .
\ee

Killing vectors that are parallel to the foliation of constant $\rho$ surfaces generate the subgroup of symmetries which preserve the foliation.  These constitute a subalgebra of Killing vectors of $G_{AB}$ for which $K^5=0$.  We choose a basis of this subalgebra and index the elements by $i$, 
\be 
K_i^A(X)=\begin{cases} K_i^\mu(x) & A=\mu \\ 0 & A=5\end{cases} \ ,
\ee
where we have written $K_i^\mu(x)$ for the $A=\mu$ components, since Killing's equations tell us they are independent of $\rho$.  We then extend this basis to that of the full algebra of all Killing vectors by appending a suitably chosen set of linearly independent Killing vectors with non-vanishing $K^5$.  We index these with $I$, so that $(K_i,K_I)$ is a basis of the full algebra of Killing vectors.  Killing's equation tells us that $K_I^5(x)$ is independent of $\rho$, and that $K_i^\mu(x)$ is a Killing vector of $g_{\mu\nu}$.

A general global symmetry transformation is
\be 
\delta_KX^A=a^i K_i^A(X)+a^I K_I^A(X) \ ,
\ee
where $a^i$ and $a^I$ are arbitrary constants.  From~(\ref{gaugefixsym}), the transformations on $\pi$ become
\be 
\label{specialcasesym} 
(\delta_K+\delta_{g,{\rm comp}})\pi=-a^i K_i^\mu(x)\partial_\mu\pi+a^I K_I^5(x)-a^I K_I^\mu(x,\pi)\partial_\mu\pi \ .
\ee
Thus the $K_i$ describe linearly realized symmetries, whereas the $K_I$ describe nonlinearly realized ones.  Therefore, the algebra of Killing vectors is spontaneously broken to the subalgebra of those preserving the foliation.

A general choice for the action~(\ref{gaugefixedaction}) will lead to scalar field equations for $\pi$ which are higher than second order in derivatives and may therefore propagate extra ghost degrees of freedom~\cite{Ostrogradski,deUrries:1998bi}.  
However, as pointed out in~\cite{deRham:2010eu}, there are a finite number of actions that lead to second order equations.
The possible
extensions of Einstein gravity which remain second order are given by Lovelock terms~\cite{Lovelock:1971yv} and their boundary terms.  
These terms are specific combinations of powers of the Riemann tensor which are topological (i.e. total derivatives) in some specific home dimension, but in lower dimensions have the property that equations of motions derived from them are second order.
The prescription of~\cite{deRham:2010eu} is then as follows: on the 4-dimensional brane, we may add the first two Lovelock terms, namely the cosmological constant term $\sim \sqrt{-\bar g}$ and the Einstein-Hilbert term $\sim \sqrt{-\bar g}R[\bar g]$.  (The higher Lovelock terms will be total derivatives in 4-dimensions.)  We may also add the boundary term corresponding to a bulk Einstein-Hilbert term, $\sqrt{-\bar g}K$, and the boundary term ${\cal L}_{\rm GB}$ corresponding to the Gauss-Bonnet Lovelock invariant $R^2 - 4 R_{\mu\nu} R^{\mu\nu}+ R_{\mu\nu\alpha\beta} R^{\mu\nu\alpha\beta}$ in the bulk.  The zero order cosmological constant Lovelock term in the bulk has no boundary term, although we may construct a tadpole-like term from it, and the higher order bulk Lovelock terms vanish identically.  Therefore, in total, for a 3-brane there are 
five terms that lead to second order equations for $\pi$,
\bea   
{\cal L}_1&=&\sqrt{-g}\int^\pi d\pi' f(\pi')^4,\nn\\
{\cal L}_2&=&- \sqrt{-\bar g} \ ,\nn\\
{\cal L}_3&=& \sqrt{-\bar g}K \ ,\nn\\
{\cal L}_4&=& -\sqrt{-\bar g}\bar R \ ,\nn\\
{\cal L}_5&=&{3\over 2}\sqrt{-\bar g} {\cal K}_{\rm GB} \ ,
\label{ghostfreegenterms} 
\eea
where the explicit form of the Gauss-Bonnet boundary term is
${\cal K}_{\rm GB}=-{1\over3}K^3+K_{\mu\nu}^2K-{2\over 3}K_{\mu\nu}^3-2\left(\bar R_{\mu\nu}-\half \bar R \bar g_{\mu\nu}\right)K^{\mu\nu}$. 

$\mathcal L_1$ is the zero derivative tadpole term which is the proper volume between any $\rho=$ constant surface and the brane position, $\pi(x)$.  While different in origin from the other terms, it too has the symmetry~(\ref{gaugefixsym}).  Each of these terms may appear in a general Lagrangian with an arbitrary coefficient. 

Evaluating these expressions for the metric~(\ref{metricform}) involves a lengthy calculation. The resulting first four terms are
\bea   
{\cal L}_1&=&\sqrt{-g}\int^\pi d\pi' f(\pi')^4,\nn\\
{\cal L}_2&=&-\sqrt{-g}f^4\sqrt{1+{1\over f^2}(\partial\pi)^2},\nn\\
{\cal L}_3&=&\sqrt{-g}\left[f^3f'(5-\gamma^2)-f^2[\Pi]+\gamma^2[\pi^3]\right],\nn \\
{\cal L}_4&=& -\sqrt{-g}\Big\{{1\over\gamma}f^2R-2{\gamma}R_{\mu\nu}\nabla^\mu\pi\nabla^\nu\pi \nn\\
&&+\gamma\left[[\Pi]^2-[\Pi^2]+2{\gamma^2\over f^2}\left(-[\Pi][\pi^3]+[\pi^4]\right)\right]\nn\\
&&\!+6{f^3f''\over \gamma}\left(-1+\gamma^2\right) \!	 \nn \\
&&+2\gamma ff'\left[-4[\Pi]+{\gamma^2\over f^2}\left(f^2[\Pi]+4[\pi^3]\right)\right]\nn\\
&&-6{f^2f'^2\over \gamma}\left(1-2\gamma^2+\gamma^4\right) \Big\} \ ,
\eea
where we refer the reader to~\cite{Goon:2011qf} for the explicit and long expression for ${\cal L}_5$.

In these expressions, all curvatures and covariant derivatives are those of the background metric $g_{\mu\nu}$.  We use the notation $\Pi$ for the matrix of derivatives $\Pi_{\mu\nu}\equiv\nabla_{\mu}\nabla_\nu\pi$.  Its traces are written as $[\Pi^n]\equiv Tr(\Pi^n)$, e.g. $[\Pi]=\nabla_\mu\nabla^\mu\pi$ and $[\Pi^2]=\nabla_\mu\nabla_\nu\pi\nabla^\mu\nabla^\nu\pi$, where all indices are raised with respect to $g^{\mu\nu}$.  We also define $[\pi^n]\equiv \nabla\pi\cdot\Pi^{n-2}\cdot\nabla\pi$, e.g. $[\pi^2]=\nabla_\mu\pi\nabla^\mu\pi$ and $[\pi^3]=\nabla_\mu\pi\nabla^\mu\nabla^\nu\pi\nabla_\nu\pi$.  We stress that no integrations by parts have been performed in obtaining these terms.  The equations of motion for these terms contain no more than two derivatives on each field, ensuring that no extra degrees of freedom propagate around any background. Note that, after integrations by parts, these actions will display the general structure presented in~\cite{Deffayet:2011gz}. In our construction, one can immediately identify the nonlinear symmetries from the bulk isometries.

\begin{figure} 
   \centering
   \includegraphics[width=3.3in]{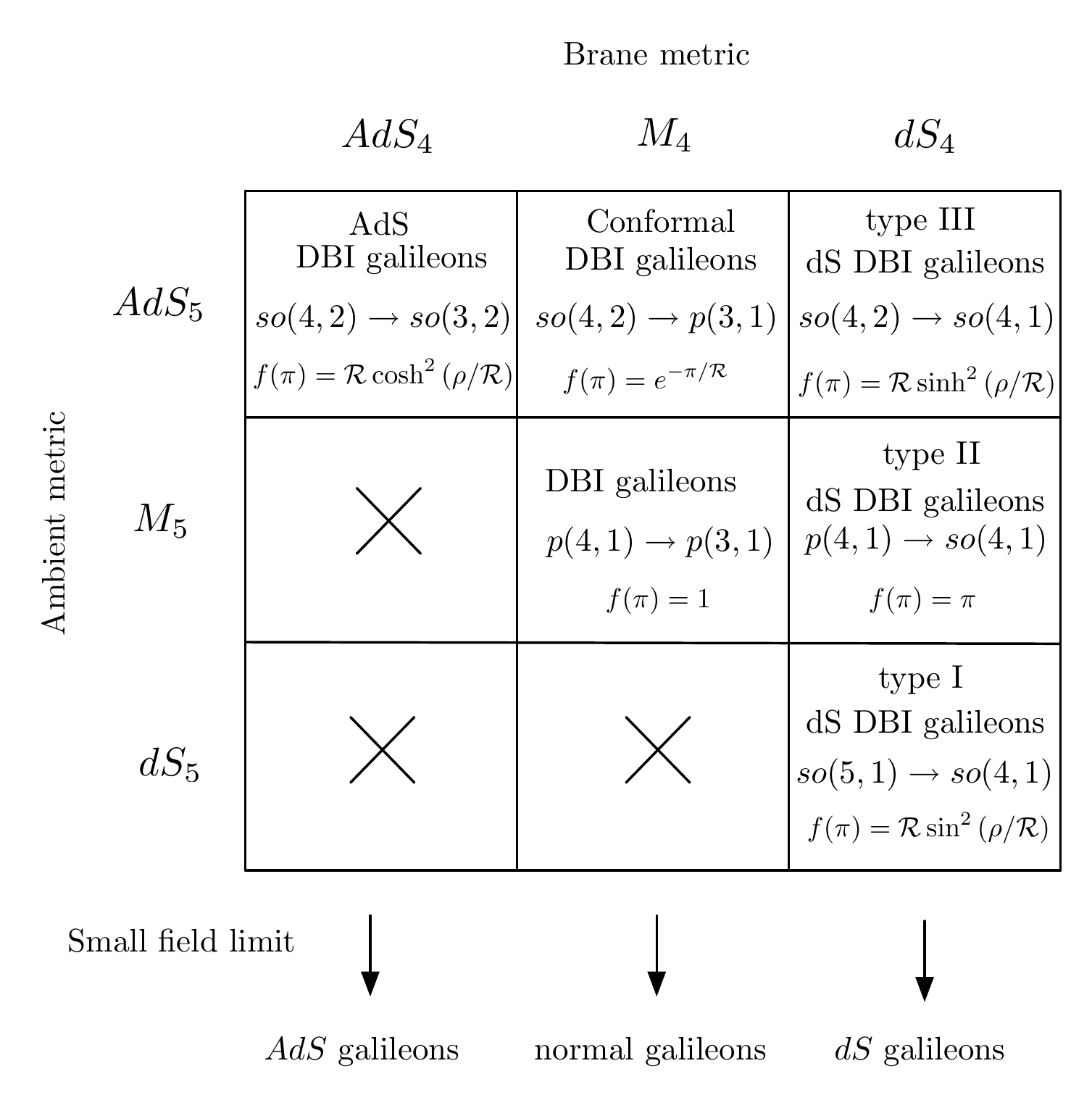}
   \caption{Types of maximally symmetric embedded brane effective field theories, their symmetry breaking patterns, and functions 
   $f(\pi)$.  The relationships to the Galileon and DBI theories are also noted~\cite{Goon:2011qf}.}
   \label{types}
\end{figure}

As shown in~\cite{Goon:2011qf} interesting examples of this construction arise when both the bulk and the leaves of the foliation are maximally symmetric, and the bulk metric has a single time component.  There are only six possibilities: Flat $M_5$ can be foliated by flat $M_4$ slices or by $dS_4$ slices; $dS_5$ can be foliated by flat $M_4$ slices, $dS_4$ slices, or $AdS_4$ slices; and $AdS_5$ can only be foliated by $AdS_4$ slices.  Each of these theories has 15 global symmetries, 10 of which are linearly realized, and 5 of which are non-linearly realized.  Each possibility has a different symmetry breaking pattern and a different value for $f(\pi)$, see Figure~\ref{types}.  Note that the only invariant data necessary for constructing a brane theory are the background metric and the action.  Theories with the same background metric and the same action are isomorphic, regardless of the choice of foliation (which is, after all, merely a choice of gauge).  

To develop the analogues of the original Galileon theory, we expand the Lagrangians in powers of $\lambda$ around some constant background, $\pi\rightarrow\pi_0+\lambda\pi$.  One can find appropriate linear combinations of the Lagrangians, $\bar{\mathcal L}_n=c_1\mathcal L_1+\ldots +c_n\mathcal L_n$, for which all terms $\mathcal{O}\left (\lambda^{n-1}\right )$ or lower are total derivatives.  This was performed in~\cite{deRham:2010eu} for the $M_4$ in $M_5$ case and the results precisely reproduce the Galileon theory.

When this prescription is carried out for the remaining four maximally symmetric cases in which the 4d background is curved, new classes of theories are produced.  After canonical normalization, $\hat\pi={1\over L^2}\pi$, where $L$ is the $dS_4$ or $AdS_4$ radius, the Lagrangians become
\begin{eqnarray} 
\hat{\cal L}_1&=&\sqrt{-g}\hat\pi \ , \nn\\
\hat{\cal L}_2&=&-\half\sqrt{-g} \left((\partial\hat\pi)^2-{R\over 3}\hat \pi^2\right) \ ,\nn \\
\hat{\cal L}_3&=& \sqrt{-g}\left(-{1\over 2}(\partial\hat\pi)^2[\hat\Pi]-{R\over 4} (\partial\hat\pi)^2\hat\pi+{R^2\over 36}\hat\pi^3\right) \ ,\nn\\
\hat{\cal L}_4&=&\sqrt{-g}\Big[-\half(\partial\hat\pi)^2\Big([\hat\Pi]^2-[\hat\Pi^2]+{R\over 24}(\partial\hat\pi)^2\nn\\
&&+{R\over 2}\hat\pi[\hat\Pi]+{R^2\over 8}\hat\pi^2\Big)+{R^3\over 288}\hat\pi^4\Big] \ , 
\label{singlesetGalileons} 
\eea 
where we have again presented only up to $n=4$ to avoid unwieldy expressions (see~\cite{Goon:2011qf} for ${\cal L}_5$).  Here $R=\pm{12\over L^2}$ is the Ricci curvature of the $dS_4$ or $AdS_4$ background. This expression also incorporates the $M_4$ background cases in which $R=0$.  These simpler Lagrangians are Galileons that live on curved space yet retain the same number of symmetries as the full theory, whose form comes from expanding~(\ref{gaugefixsym}) in appropriate powers of $\lambda$.  In the case of a $dS_4$ background in conformal inflationary coordinates $(u,y^i)$, the non-linear symmetries are
\be 
\label{dSGalileontrans}
\delta_{+}\hat\pi={1\over u}\left(u^2-y^2\right) , \ \ \
\delta_{-} \hat\pi=-{1\over u},\ \ \ 
\delta_{i} \hat\pi = {y_i\over u} \ .
\ee
	
A striking feature of these fully covariant models, which is not present in the flat space Galileon theories, is the presence of potentials with couplings determined by the symmetries~(\ref{gaugefixsym}).  In particular, the scalar field acquires a mass of order the $dS_4$ or $AdS_4$ radius, protected against renormalization by the symmetries~(\ref{dSGalileontrans}).

In the $dS_4$ case the field is either a ghost or has a tachyonic mass.  While ghosts are to be avoided, a tachyonic mass need not be a disaster, since the higher order Lagrangians can be added to stabilize the field.  For either the $dS_4$ or $AdS_4$ case, one might impose a $Z_2$ symmetry so that the full Lagrangian is of the form $\hat{ \mathcal{L}}=\pm\left (\hat{\mathcal L}_2-a\hat{\mathcal L}_4\right )$.  Choosing the appropriate overall sign ($+$ for $dS_4$ and $-$ for $AdS_4$) and $a>0$, allows for spontaneous symmetry breaking with the $Z_2$ preserving unstable vacuum at $\pi=0$, and the $Z_2$ breaking stable vacua at $\pi=\pm\sqrt {24\over a}{1\over |R|}$.  None of the vacua alter the symmetries of the theory and re-expansion of the fields around the spontaneously broken vacua gives a combination of the Lagrangians~(\ref{singlesetGalileons}) whose relative coefficients depend only upon the constant $a$.  Explicitly, expansion around the positive vacua gives
\bea 
-2\hat{\cal L}_2-\sqrt{6a} \hat{\cal L}_3-a \hat{\cal L}_4,\ \ \ dS \ ,\nn\\ 
2 \hat{\cal L}_2-\sqrt{6a} \hat{\cal L}_3+a \hat{\cal L}_4.\ \ \ AdS \ .
\eea
Thus, the resulting excitation is a ghost in the $dS_4$ case, but has healthy kinetic term in the $AdS_4$ case. This is the opposite of the results obtained by expanding around the $\pi=0$ vacuum.

In summary, we have constructed effective field theories that extend the Galileon theories to curved space in a way that explicitly identifies when there are non-linear symmetries, and have commented on the maximally symmetric examples.  These theories exhibit distinctive new features, such as the existence of potentials, with masses fixed by symmetries, opening up the possibility of new natural implementations of accelerating cosmological solutions in theories naturally having a de Sitter solution. We are currently exploring the phenomenology of these theories, including the cosmological implications, the existence of topological defects, and stability around nontrivial gravitational backgrounds.

\acknowledgments

This work is supported in part by NASA ATP grant NNX08AH27G, NSF grant PHY-0930521, and by Department of Energy grant DE-FG05-95ER40893-A020.

\bibliography{general-letter-7}

\end{document}